\documentclass{iau} 
\usepackage{graphicx}

\title[Low-Gamma Jets from Compact Binary Mergers] 
{Low-$\Gamma$ jets from Compact Binary Mergers as Candidate Electromagnetic Counterparts to Gravitational Wave Sources}

\author[Gavin P. Lamb \& Shiho Kobayashi]   
{Gavin P. Lamb
 \and Shiho Kobayashi}

\affiliation{Astrophysics Research Institute, Liverpool John Moores University, L3 5RF, UK \\ email: {\tt g.p.lamb@2010.ljmu.ac.uk}
}

\pubyear{2017}
\volume{324}  
\jname{New Frontiers in Black Hole Astrophysics}
\editors{Andreja Gomboc}
\begin{document}

\maketitle

\begin{abstract}
Compact binary mergers, with neutron stars or neutron star and black-hole components, are thought to produce various electromagnetic counterparts: short gamma-ray bursts (GRBs) from ultra-relativistic jets followed by broadband afterglow; semi-isotropic kilonova from radioactive decay of r-process elements; and late time radio flares; etc. If the jets from such mergers follow a similar power-law distribution of Lorentz factors as other astrophysical jets then the population of merger jets will be dominated by low-$\Gamma$ values. The prompt gamma-rays associated with short GRBs would be suppressed for a low-$\Gamma$ jet and the jet energy will be released as X-ray/optical/radio transients when a shock forms in the ambient medium. Using Monte Carlo simulations, we study the properties of such transients as candidate electromagnetic counterparts to gravitational wave sources detectable by LIGO/Virgo. Approximately 78\% of merger-jets result in failed GRB with optical peaks 14-22 magnitude and an all-sky rate of 2-3 per year.
\end{abstract}

\firstsection 
\section{Introduction}

The merger of a binary system due to gravitational wave (GW) emission where the binary components are either neutron stars (NS) or stellar mass black holes (BH) is a potential source of GWs detectable by current gravitational wave detectors (e.g. LIGO/Virgo).
In 2015, advanced LIGO made two unambiguous detections at 5-$\sigma$ (GW150914 and GW151226) and a third possible detection at 87$\%$ confidence (LVT151012) of the inspiral and merger of binary stellar mass BH systems \cite[(The LIGO Scientific Collaboration et al. 2016)]{2016arXiv160604856T}.
An electromagnetic (EM) counterpart to the inspiral and merger of a binary BH system is not generally expected and no EM counterpart was detected for any of the observed GW signals \cite[(Copperwheat et al. 2016)]{2016MNRAS.462.3528C}.

The next GW breakthrough will be the detection of the inspiral and merger of NS-BH or NS-NS systems.
For such mergers we expect various EM counterparts, where the nature of the counterpart depends on the viewing angle \cite[(Metzger \& Berger 2012)]{2012ApJ...746...48M}.
When a NS-BH or NS-NS system merge, the merger ejecta forms an accretion disk, tidal tail and disk wind.
The rapid accretion from a disk onto a newly formed BH is thought to power bipolar ultrarelativistic jets responsible for short GRBs and their afterglow, the typical timescale for a short GRB is seconds and hours to days for the associated afterglow (\cite[Woosley \& Bloom 2006]{2006ARA&A..44..507W}; \cite[Nakar 2007]{2007PhR...442..166N}; \cite[Berger 2014]{2014ARA&A..52...43B}).
The more isotropic ejecta can power r-process nucleosynthesis where radioactive decay gives rise to macronova emission, these occur on a day to weeks timescale (\cite[Tanvir et al. 2013]{2013Natur.500..547T}; \cite[Berger et al. 2013]{2013ApJ...774L..23B}; \cite[Tanaka 2016]{2016AdAst2016E...8T}).
At later times, approximately a month, the merger ejecta can interact with the ambient medium giving rise to radio flares (\cite[Nakar \& Piran 2011]{2011Natur.478...82N}; \cite[Hotokezaka et al. 2016]{2016arXiv160509395H}).

\section{Gamma-ray Bursts and Failed GRBs}

Except for two cases (\cite[Cenko et al. 2013]{2013ApJ...769..130C}, \cite[2015]{2015ApJ...803L..24C}) we currently detect GRBs and their afterglow by a high energy $\gamma$-ray trigger.
The variability timescale of the prompt $\gamma$-ray emission indicates a dependence for the emiting region, $R_d \propto \Gamma^2$, where $R_d$ is the dissipation radius and $\Gamma$ the bulk Lorentz factor.
The photospheric radius $R_\star$, the point at which the fireball becomes optically thin, can be conservatively estimated by considering the density of electrons associated with a baryonic outflow, $R_\star \propto E^{1/2}\Gamma^{-1/2}$.
These conditions require GRBs to be produced in ultrarelativistic jets with $\Gamma \sim 100$.

For jets with $\Gamma << 100$ the dissipation radius falls below the photoshere and the prompt $\gamma$-rays would be suppressed.
Such a failed GRB will still have a broadband orphan afterglow.
The observation of afterglows associated with the relativistic jets from binary NS mergers is currently biased by the requirement of a high energy $\gamma$-ray trigger, GW triggered searches may reveal a hidden population of low-$\Gamma$ merger jets by their on-axis orphan afterglow.
Astrophysical jets from other accreting BH systems (e.g. AGN, blazars) follow a simple power law distribution with a negative index, $N(\Gamma) \propto \Gamma^{-a}$, where $a$ is $\sim2$ for blazars \cite[(Saikia et al. 2016)]{2016MNRAS.461..297S}.
If the jets from mergers follow a similar distribution then failed GRBs would outnumber short GRBs.

\begin{figure}[h!t]
\begin{center}
 \includegraphics[width=3.4in]{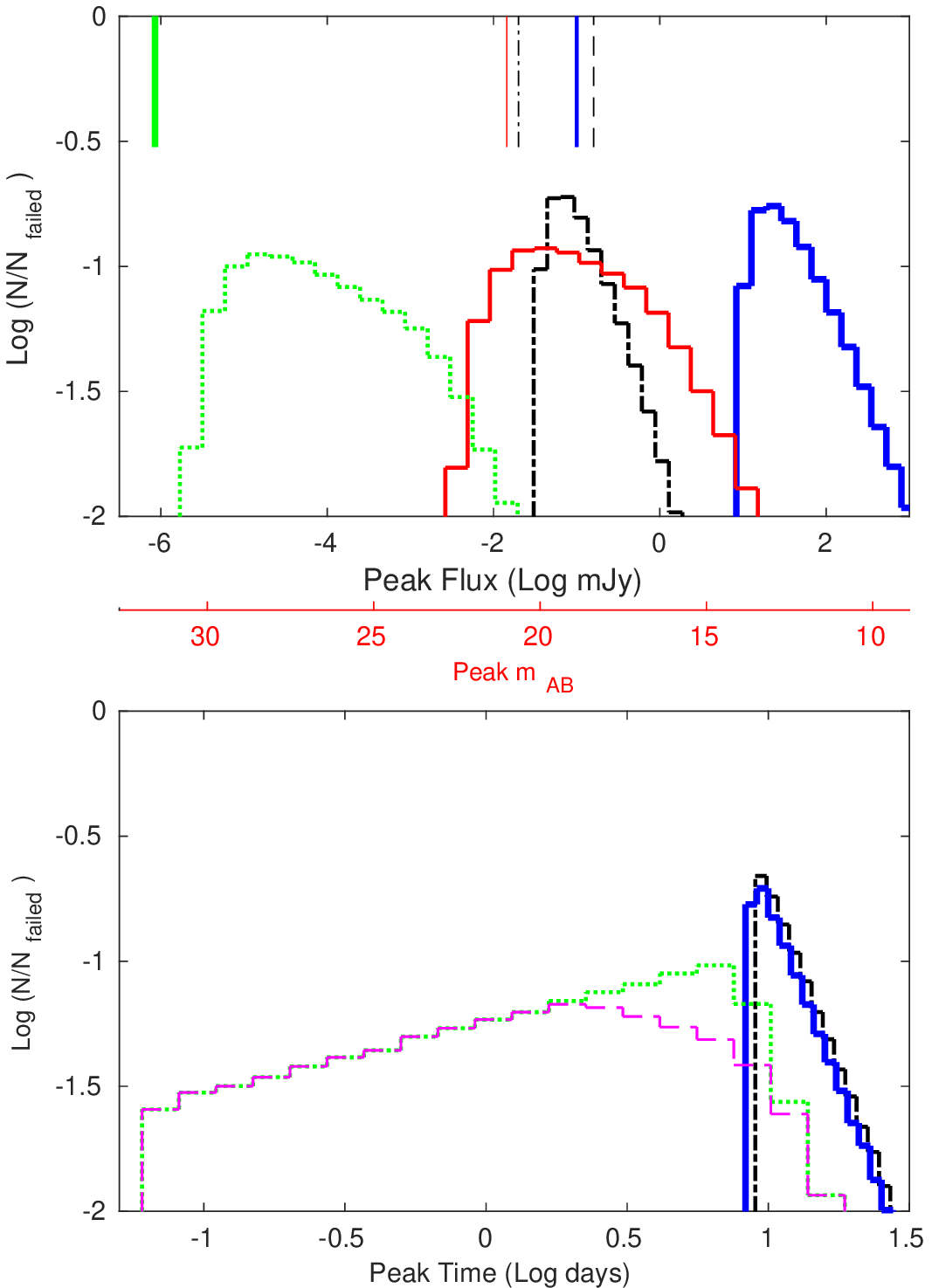} 
 \caption{The peak-flux (top panel) and peak-time (bottom panel) distribution of 
on-axis orphan afterglow from failed GRB events within 300 Mpc. 
The distributions are normalized by the total number of failed GRBs. 
X-ray (dotted green line), optical (thick solid red line), radio 10 GHz (thick solid blue line) and 
radio 150 MHz (thick dash-dotted black line).   
The vertical lines in the top panel indicate the sensitivity limits of telescopes (thick green XRT, thin red optical $\sim 2$ m, dash-dotted SKA1-Low, and dashed 48 LOFAR), and 
the dashed magenta line in the bottom panel shows the distribution of
bright events $m_g \leq 21$ \cite[(Lamb \& Kobayashi 2016)]{2016ApJ...829..112L}.}
   \label{f1}
\end{center}
\end{figure}

\section{Monte Carlo}

To test whether the on-axis orphan afterglow of a failed GRB from a compact stellar merger would be observable, given a GW detection, a Monte Carlo of $2\times 10^5$ events was generated.
The events followed the \cite[Wanderman \& Piran (2015)]{2015MNRAS.448.3026W} redshift and luminosity distribution for short GRBs.
The bulk Lorentz factor followed a simple power law with index $a=1.75$ in the limits $3\leq \Gamma \leq 10^3$, and independent of a bursts energy.
For details of the Monte Carlo and radiative transfer see \cite[Lamb \& Kobayashi (2016)]{2016ApJ...829..112L}.

\section{Results}

We found that for a sample in the range $<300$ Mpc, the LIGO/Virgo detection limit for face-on NS-NS mergers, that events below a line given by $\Gamma \sim 16(E_{\rm K}/10^{50}{\rm erg})^{0.15}$ always result in failed GRBs, where $E_{\rm K}$ is the isotropic equivalent kinetic energy.
For the parameters used we found that 78$\%$ of mergers resulted in failed GRBs; the peak flux and time for the on-axis orphan afterglows from the failed GRBs is shown in figure \ref{f1}.
The model parameters used for the afterglows are: 
$n=10^{-1}$ protons cm$^{-3}$, microphysical parameters $\epsilon_B=10^{-2}$ and $\epsilon_e=10^{-1}$, the index of the power-law distribution of random electrons accelerated at shock $p=2.5$, and the jet half-opening angle $\theta_j=20^{\circ}$ ensuring that the jet break time is later than the deceleration time for our sample and is within the limits $16 \pm 10^{\circ}$ found by \cite[Fong et al. (2015)]{2015ApJ...815..102F} for short GRBs.
The jet opening angle plays a role only when we estimate the jet break time.

\begin{figure}[h!t]
\begin{center}
\includegraphics[width=3.4in]{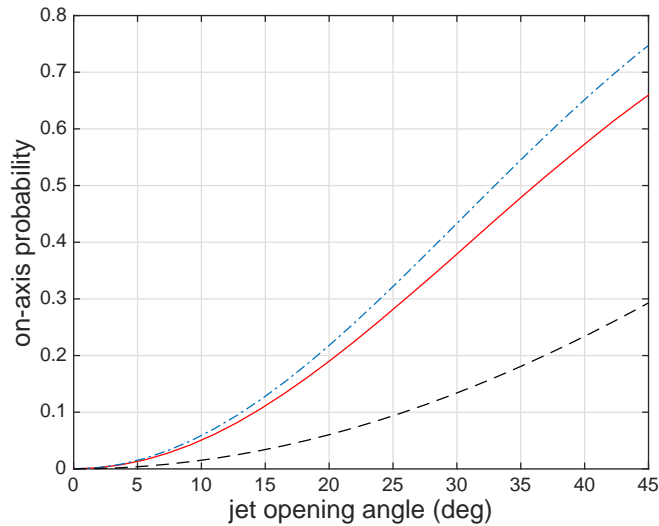} 
\caption{On-axis probability as a function of a jet half-opening angle $\theta_j$.
The beaming factor $f_b=1-\cos\theta_j$ (black dashed line), given a GW detection the simple approximation 
$A^3f_b$ (blue dash-dot line), and the numerical results (red solid line).}
\label{f2}
\end{center}
\end{figure}

\section{Discussion \& Conclusion}

GW emission is strongest on-axis from a merger event \cite[(Kochanek \& Piran 1993)]{1993ApJ...417L..17K}, thus the probability of an on-axis merger is higher, given a GW detection, than if we consider only the isotropic case (see figure \ref{f2}).
If the Lorentz factor of a jet is correlated with the opening angle, where lower $\Gamma$ gives a wider half-opening angle (as indicated for long GRBs by \cite[Ghirlanda et al. 2013]{2013MNRAS.428.1410G}), then the rates of on-axis orphan afterglow following a GW trigger will be higher than those given here.
By using the rate of short GRBs within the LIGO detection volume predicted from the Swift short GRB rate by \cite[Metzger \& Berger (2012)]{2012ApJ...746...48M}, 0.03 per year, and considering the all sky rate - we find approximately 2.6(26) on-axis orphan afterglows from NS-NS(NS-BH) mergers per year when assuming a jet half-opening angle of $20^\circ$.

EM counterparts to GW emission from NS-NS and NS-BH mergers include radio flares, macronova, short GRBs, and on and off axis (orphan) afterglow.
A GW triggered search for EM counterparts could reveal a hidden population of low-$\Gamma$ merger jets from such events and on-axis orphan afterglow of failed GRBs are therefore a strong candidate for EM follow-up searches.
The (non)detection of such sources will help constrain the Lorentz factor distribution of such jets (e.g. clustered at high-$\Gamma$, a power-law, a log-normal, or multiple populations), and provide constraints on the acceleration process of relativistic jets.

This research was supported by STFC grants. 
GPL was supported by an IAU travel grant.

\end{document}